\RequirePackage{fix-cm}
\documentclass[twocolumn]{svjour3-arxiv}          % twocolumn
\smartqed  % flush right qed marks, e.g. at end of proof
\usepackage{color,soul}

\usepackage{graphicx}
\usepackage{subcaption}
\usepackage[]{hyperref}
\hypersetup{%
    colorlinks=true,%
    citecolor=blue,%
    linkcolor=blue,%
    urlcolor=blue,%
}
 \journalname{Computing and Software for Big Science}
\begin{document}
\sloppy %important for not going outside boundaries in the columns
\title{MLaaS4HEP: Machine Learning as a Service for HEP}

\author{
Valentin Kuznetsov \and
Luca Giommi \and
Daniele Bonacorsi
}

%\authorrunning{Short form of author list} % if too long for running head

\institute{Valentin Kuznetsov \at
              Cornell University, Ithaca, USA \\
              \email{vkuznet@gmail.com} \\
              ORCiD: 0000-0003-0667-069X
           \and
           Luca Giommi \at University of Bologna and INFN, Bologna, Italy \\
           \email{luca.giommi3@unibo.it}\\
           ORCiD: 0000-0003-3539-4313
            \and
           Daniele Bonacorsi \at University of Bologna and INFN, Bologna, Italy \\
           \email{daniele.bonacorsi@unibo.it}\\
           ORCiD: 0000-0002-0835-9574
}

%\date{Received: date / Accepted: date}
%
\maketitle

\begin{abstract}
Machine Learning (ML) will play a significant role in the success of the
upcoming High-Luminosity LHC (HL-LHC) program at CERN. An unprecedented
amount of data at the exascale will be collected by LHC experiments
in the next decade, and this effort will require novel approaches to train
and use ML models.
In this paper, we discuss a Machine Learning as a
Service pipeline for HEP (MLaaS4HEP) which provides three independent layers: a
data streaming layer to read High-Energy Physics (HEP) data in their native
ROOT data format; a data training layer to train ML models using
distributed ROOT files; a data inference layer to serve predictions using pre-trained ML models
via HTTP protocol.
Such modular design opens up the possibility to train
data at large scale by reading ROOT files from remote storage facilities, e.g.
World-Wide LHC Computing Grid (WLCG) infrastructure, and feed the data to the user's
favorite ML framework.
The inference layer implemented as TensorFlow as a Service (TFaaS)
may provide an easy access to pre-trained ML models in
existing infrastructure and applications inside or outside of the HEP domain.
In particular, we demonstrate the usage of the MLaaS4HEP architecture for a
physics use-case, namely the $t\bar{t}$ Higgs analysis in CMS
originally performed using custom made Ntuples. We provide details on the
training of the ML model using distributed ROOT files, discuss the performance
of the MLaaS and TFaaS approaches for the selected physics analysis, and compare the
results with traditional methods.

% \PACS{PACS code1 \and PACS code2 \and more}
% \subclass{MSC code1 \and MSC code2 \and more}
\end{abstract}

\section{Introduction}

With the CERN LHC program underway, we started seeing an exponential
acceleration of data growth in the HEP field.  By the end of Run II, the CERN
experiments were already operating at the Peta-Byte (PB) level, producing
$O$(100) PB of data each year. The new HL-LHC program will extend it further, to
the Exa-Byte scale, and the usage of ML in HEP will be critical \cite{MLCWP}. ML techniques have been
successfully used in online and offline reconstruction programs, and there is a
huge gain in applying them to detector simulation, object reconstruction,
identification, Monte-Carlo (MC) generation, and beyond \cite{HEPML}.  As was
pointed out in the ML in HEP Community White Paper \cite{MLCWP} the lack of engagement from
Computer Science experts to address HEP ML challenges is partly due to the fact that HEP
data are stored in ROOT data-format, which is mostly unknown outside
of the HEP community. 
Moreover, the existing ML frameworks rely on fixed-size data representation of
individual events, usually stored in CSV \cite{csv}, NumPy \cite{NumPy}, HDF5
\cite{hdf5} data formats, while in HEP the size of individual events cannot be
determined a-priory\footnote{For instance, the number of electrons varies in each
physics event.}, and the data are stored in the event tree-based
data-structures used by the ROOT \cite{ROOT} data-format, and may require custom
C++ classes to decode them properly. This and other reasons\footnote{The event-based data structures cannot be fed directly to existing ML frameworks and
special care should be taken either at the framework or at the data input level
discussed in this paper.} led to an artificial gap between ML and
HEP communities.
For example, in recent Kaggle challenges \cite{kaggleATLAS,kaggletracking,flavours-of-physics}
the HEP data was presented in CSV data-format to allow non-HEP ML practitioners to compete.
Moreover, production workflows quite often require additional transformations,
e.g. in CMS a Deep Neural Network (DNN) used for a jet tagging
algorithm relies on the TensorFlow (TF) queue system with a custom operation kernel for reading
ROOT trees and feeding them to ML models like TensorFlow
\cite{existing_workflows}.
Here we discuss the Machine Learning as a Service (MLaaS) architecture for HEP, referred
to as MLaaS4HEP in this paper, which consists
of two individual parts. The first part, the MLaaS4HEP framework \cite{MLaaS4HEP},
provides a way to read HEP ROOT-based data natively into the Python ML framework
of user choice. And, the second part, the TensorFlow as a Service (TFaaS) framework \cite{TFaaS}, 
can be used to host pre-trained ML models and obtain ML predictions via HTTP protocol.

This approach can be used by physicists or experts outside of HEP domain because
it only relies on Python libraries. It provides access to local or remote data storage,
and does not require any modification or
integration with the experiment's specific framework(s).  Such modular design opens
up a possibility to train ML models on PB-size datasets remotely accessible
from the WLCG sites without requiring data transformation and data locality.
Therefore, an existing gap between HEP and ML communities can be easily closed
using the discussed MLaaS architecture.

The organization of this paper is the following.
Section \ref{related_works} provides a summary of related works and the key
aspects of the proposed solution. Section \ref{Architecture} presents the
details of the MLaaS4HEP architecture and its workflow.
Section \ref{real-case} shows performance results and validation of MLaaS4HEP for a physics
use-case. Section \ref{Improvements} summarizes possible future directions,
and Section \ref{summary} presents the summary.

\section{Related works and solutions}\label{related_works}
Machine Learning as a Service is a well-known concept in industry, and major IT
companies offer such solutions to their customers. For example, Amazon ML,
Microsoft Azure ML Studio, Google Prediction API and ML engine, and IBM Watson
are prominent implementations of this concept (see \cite{MLaaScomparison}).
Usually, Machine Learning as a Service is used as an umbrella of various ML
tasks such as data pre-processing, model training and evaluation, and inference
through REST APIs. Even though providers offer plenty of interfaces and APIs, most
of the time these services are designed to cover standard use-cases, e.g.
natural language processing, image classifications, computer
vision, and speech recognition. Even though a custom ML codebase can be supplied
to these platforms, its usage for HEP is quite limited for several reasons.
For instance, the HEP ROOT data-format cannot be used directly in any 
service provider's APIs. Therefore, the operational cost, e.g. data transformation from ROOT
files to data-format used by MLaaS provider APIs, data management, and data
pre-processing, can be very significant for large datasets. The data flattening
from dynamic size event-based tree format to fixed-size data representation
does not exist. Therefore, we found that out-of-the-box commercial solutions
most often are not applicable or ineffective for HEP use-cases (cost-wise and
functionality-wise).
This might change in the future, as various initiatives,
e.g. CERN OpenLab \cite{CERNOpenLab}, continue to work in close cooperation with almost
all aforementioned service providers.

At the same time, various R\&D activities within HEP are underway.  For
example, the hls4ml project \cite{hls4ml} targets ML inference on FPGAs, while
the SonicCMS project \cite{SonicCMS} is designed as Services for Optimal
Network Inference on Co-processors.  Both are targeted to the
optimization of the inference phase rather than  the whole ML pipeline, i.e.
from reading data to training models and serving predictions. Another 
solution uses the Spark platform for data processing and ML training \cite{CERNBigML}.
Although it seems very promising, it requires data ingestion
into the CERN EOS filesystem or the HDFS/Spark infrastructure. As such, there is no
easy way to access data located at WLCG sites or from outside of such dedicated
infrastructure. Besides, a Spark-based library (Analytics Zoo, BigDL) may be
required on top of Keras API, and flexibility of ML framework choice is limited
on the user side.
In the end, we found that there is no final product that can be
used as Machine Learning as a Service for distributed HEP data without
additional efforts which can provide transparent integration with existing Python-based
ML frameworks to perform ML training over HEP data, and this work aims to close this gap.

\subsection{Novelty of the proposed solution\label{novelty}}
The novelty of the proposed solution is the following.
\begin{enumerate}
    \item 
        We provide a transparent access to HEP datasets stored in the event
        tree-based ROOT data-format into existing Python-based ML frameworks of
        user's choice. Usually, they are
        designed to operate with row-based data structures like NumPy arrays,
        CSV files and alike. The proposed solution discussed in Sections
        \ref{Streaming} and \ref{Training} relies on the uproot library
        \cite{uproot} and XrootD protocol \cite{XrootD} for reading tree-based
        ROOT files from local filesystem or remote sites. It transforms the Jagged
        Arrays\footnote{Jagged Array is an array of arrays of which the member
        arrays can be of different sizes, see Sect. \ref{Training} for more details.}
        representation of ROOT data, and fed it into ML framework via vector or matrix-based
        transformations applied to the I/O stream. This opens up a possibility to
        use favorite ML frameworks like PyTorch \cite{PyTorch},
        TensorFlow \cite{TF}, fast.ai \cite{fastai}, etc., and train
        ML models using distributed HEP datasets.
    \item 
        We demonstrate in Sections \ref{validation} and \ref{performance}
        that the proposed solution can work at any scale and transparent to data locality.
        For that, we compared traditional HEP analysis based on custom flat tuples (derived
        from the production ROOT files) with the MLaaS4HEP approach.
        We show that the latter has several advantages such as usage of non-HEP ML frameworks,
        ability to work with local or remote storage, small or large datasets, and
        does not require domain knowledge and HEP software infrastructure.
    \item
        We provide an independent Tensor as a Service framework \cite{TFaaS}, developed as a
        part of this work, which provides access to any kind of Tensor-based ML models
        via HTTP protocol. Even though similar solutions exist in the business
        world, most of them are integrated as a part of their service stack which
        may not be affordable or accessible to research communities where an
        efficient, scalable open-source alternative is desired to have. For
        instance, the TFaaS service can be used with any programming
        language, frameworks, and scripts without additional development and
        modifications in existing infrastructure(s). It can be part of the MLaaS4HEP
        pipeline or used independently to serve predictions from non-HEP ML models.
    \item Finally, we demonstrate that the proposed architecture can be
        easily adapted among any HEP experiment either as an entire pipeline
        or be used partially. For example, the Data Streaming Layer (see Sect.
        \ref{Streaming}) provides an access to distributed ROOT files.
        The Data Training Layer (see Sect. \ref{Training}) performs proper data
        transformation from Jagged Arrays into flat data-format suitable for
        ML framework of user choice and it can be easily integrated within existing or new
        Python-based ML framework of user's choice.  The Data Inference
        Layer (see Sect. \ref{Inference}) provides the TensorFlow as a Service
        (TFaaS) service which can be used as an independent repository for pre-trained
        ML models across different experiments or independent physics analysis
        groups.
\end{enumerate}

\section{MLaaS4HEP architecture}\label{Architecture}
\begin{figure*}[!t]
\centering
\includegraphics[width=0.8\textwidth]{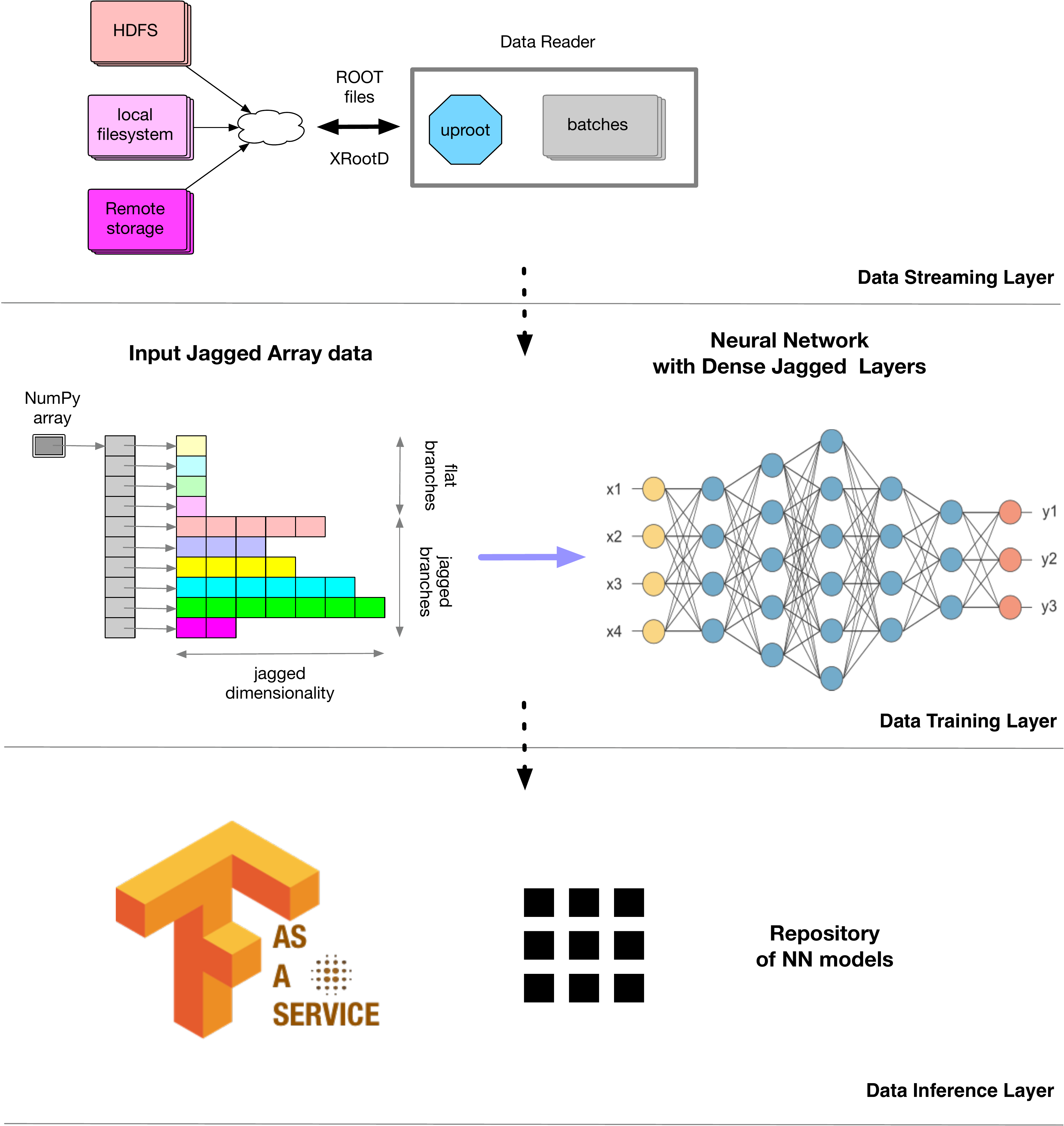}
    \caption{MLaaS4HEP architecture diagram representing three independent layers:
    a Data Streaming Layer to read local or remote ROOT files, a Data Training Layer
    to feed tree-based HEP data into ML framework, and a Data Inference Layer
    via TensorFlow as a Service \cite{MLaaS_Valentin}}
\label{fig:MLaaSArchitecture}
\end{figure*}

A typical ML workflow consists of several steps: acquire the data necessary for
training, use a ML framework to train the model, and utilize the trained model
for predictions. In our Machine Learning as a Service solution, MLaaS4HEP
\cite{MLaaS4HEP}, this workflow can be abstracted as data streaming, data
training, and inference phases, respectively. Each of these components can be either tightly
integrated into the application design, or composed and used individually.  The
choice is mostly driven by particular use cases. We can define these
layers as following (see Fig. \ref{fig:MLaaSArchitecture}).
\begin{itemize}
\item {\bf Data Streaming Layer:} it is responsible for reading local and/or remote ROOT files,
    and streaming data batches upstream to the Data Training Layer. The implementation of
    this layer requires the ROOT I/O layer with the support of remote I/O file access;
\item {\bf Data Training Layer:} it represents a thin wrapper around standard ML libraries
    such as TensorFlow, PyTorch, and others. It reads data from the Data
    Streaming Layer in chunks, transforms them from the ROOT TTree-based
    representation to the format suitable for the underlying ML framework, and
    uses it for training purposes;
\item {\bf Data Inference Layer:} it refers to the inference part of pre-trained models and
    can be either tightly integrated within the underlying HEP framework, or represented
    as a Service (aaS).
\end{itemize}
Even though the implementation of these layers can differ from one experiment
to another (or other scientific domains), it can be
easily generalized and be part of the foundation for a generic Machine Learning
as a Service framework. The MLaaS4HEP framework \cite{MLaaS4HEP} implements
the Data Streaming and Data Training layers, and we provide their details in
Sect. \ref{Streaming} and Sect. \ref{Training}, respectively. In Sect. \ref{workflow}
we provide technical details of the ML training workflow implemented in the
MLaaS4HEP framework and used for our studies presented in Sect. \ref{real-case}.
The Data Inference Layer is implemented as independent TFaaS \cite{TFaaS}
framework since it can be used outside of HEP, and its details are discussed in Sect. \ref{Inference}.

\subsection{Data Streaming Layer}\label{Streaming}
The Data Streaming Layer is responsible for streaming data from local or remote
data storage. Originally, the reading of ROOT files was mostly possible from
C++ frameworks, but the recent development of ROOT I/O now allows to easily access
ROOT data locally from Python.
The main development was done in the {\bf uproot} \cite{uproot} framework supported by the
DIANA-HEP initiative \cite{DIANAHEP}. The uproot library uses NumPy
\cite{NumPy} calls to rapidly cast data blocks in ROOT file as NumPy
arrays. It allows, among the
implemented features, a partial reading of ROOT TBranches, non-flat
TTrees, non TTrees histograms, and more. It relies on data caching and parallel
processing to achieve high throughput. In our benchmarks, we were able to read
HEP events at the level of $\sim O(10) $ kHz\footnote{Speed varies based on many factors,
including caching, type of storage and network bandwidth.} from local and from remote
storages. The latter was provided via XrootD protocol \cite{XrootD}.

In our implementation of Machine Learning as a Service (see Sect.
\ref{Prototype}) this layer was composed as a Data Generator\footnote{A piece
of code defined as Python generator to read an appropriate chunk of data upon
request from upstream code.} which is capable of reading chunk of data
either from local or remote file(s). 
The output of the Data Generator is a NumPy
array with flat and Jagged Array attributes. Such implementation provides
efficient access to large datasets since it does not require loading the entire
dataset into the RAM of the training node. Also, it can be used to parallelize the data
flow into the ML workflow pipeline. The size of data chunks read by this layer can
be easily fine-tuned based on the complexity of events and the available
bandwidth. For instance, in our initial proof-of-concept implementation, see
Sect. \ref{Prototype}, we used 1k events as a chunk data size, while within
performance studies discussed in Sect. \ref{performance} we extended chunk
size to 100k events.

\subsection{Data Training Layer}\label{Training}
\begin{figure*}[!t]
\centering
    \includegraphics[width=0.4\textwidth]{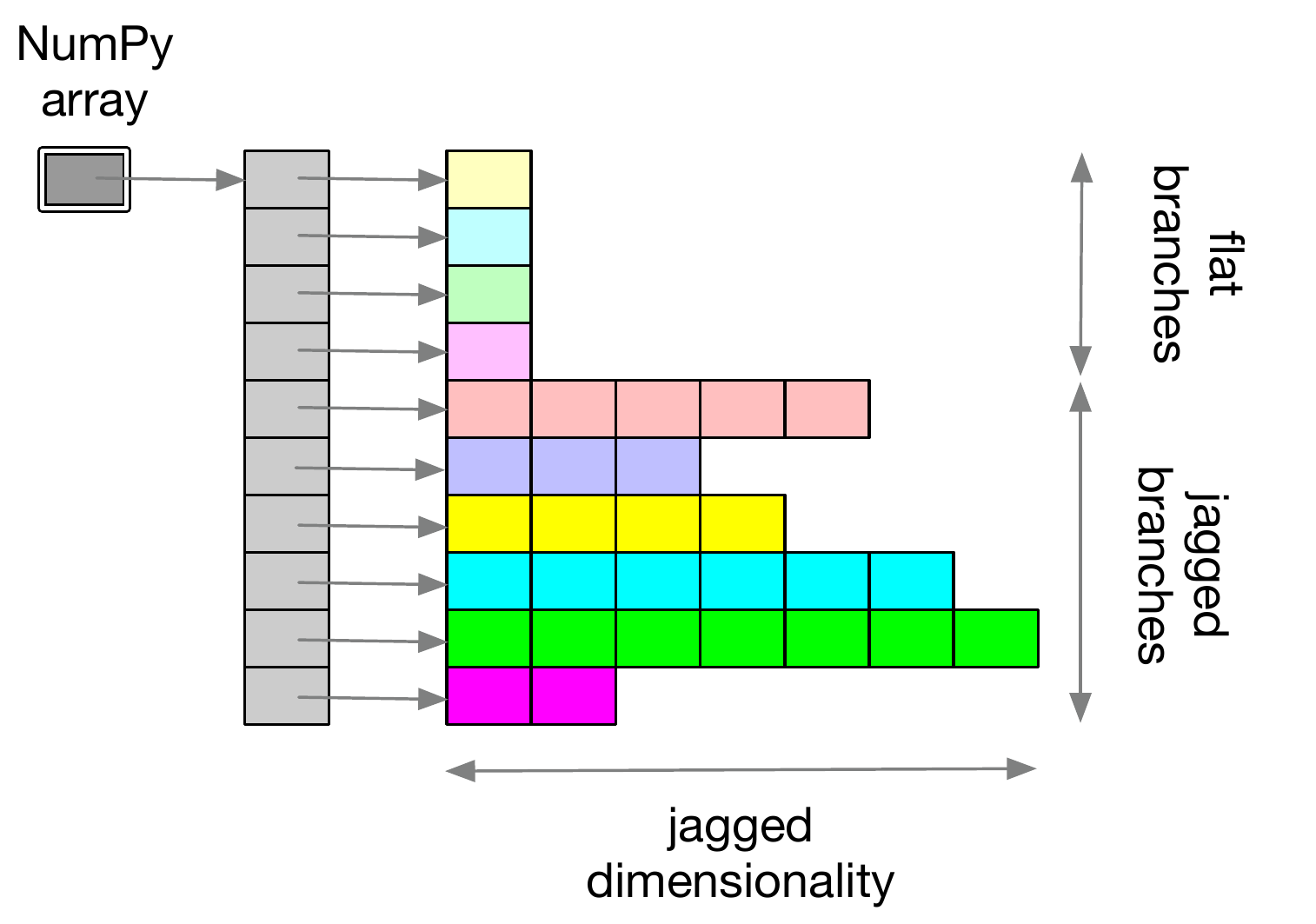}
    \caption{Jagged Array data representation. It consists of flat attributes followed by
    Jagged attributes whose dimensions vary event by event \cite{MLaaS_Valentin}}
\label{fig:JaggedArray}
\end{figure*}

\begin{figure*}[!t]
\centering
    \includegraphics[width=0.6\textwidth]{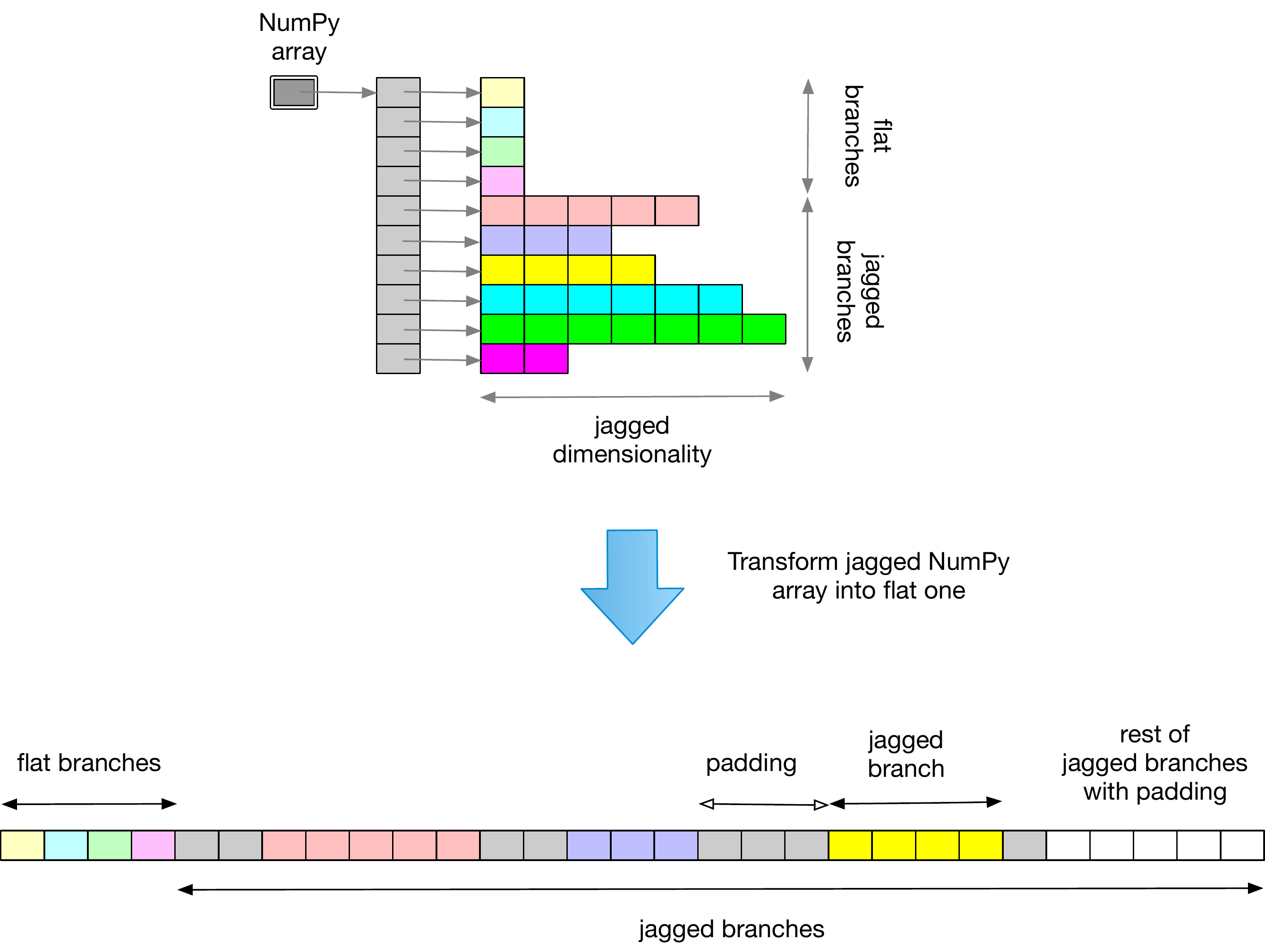}
    \caption{A vector representation of Jagged Array with padded values \cite{MLaaS_Valentin}}
\label{fig:JaggedArray2Vector}
\end{figure*}

\begin{figure*}[!t]
\centering
    \includegraphics[width=0.6\textwidth]{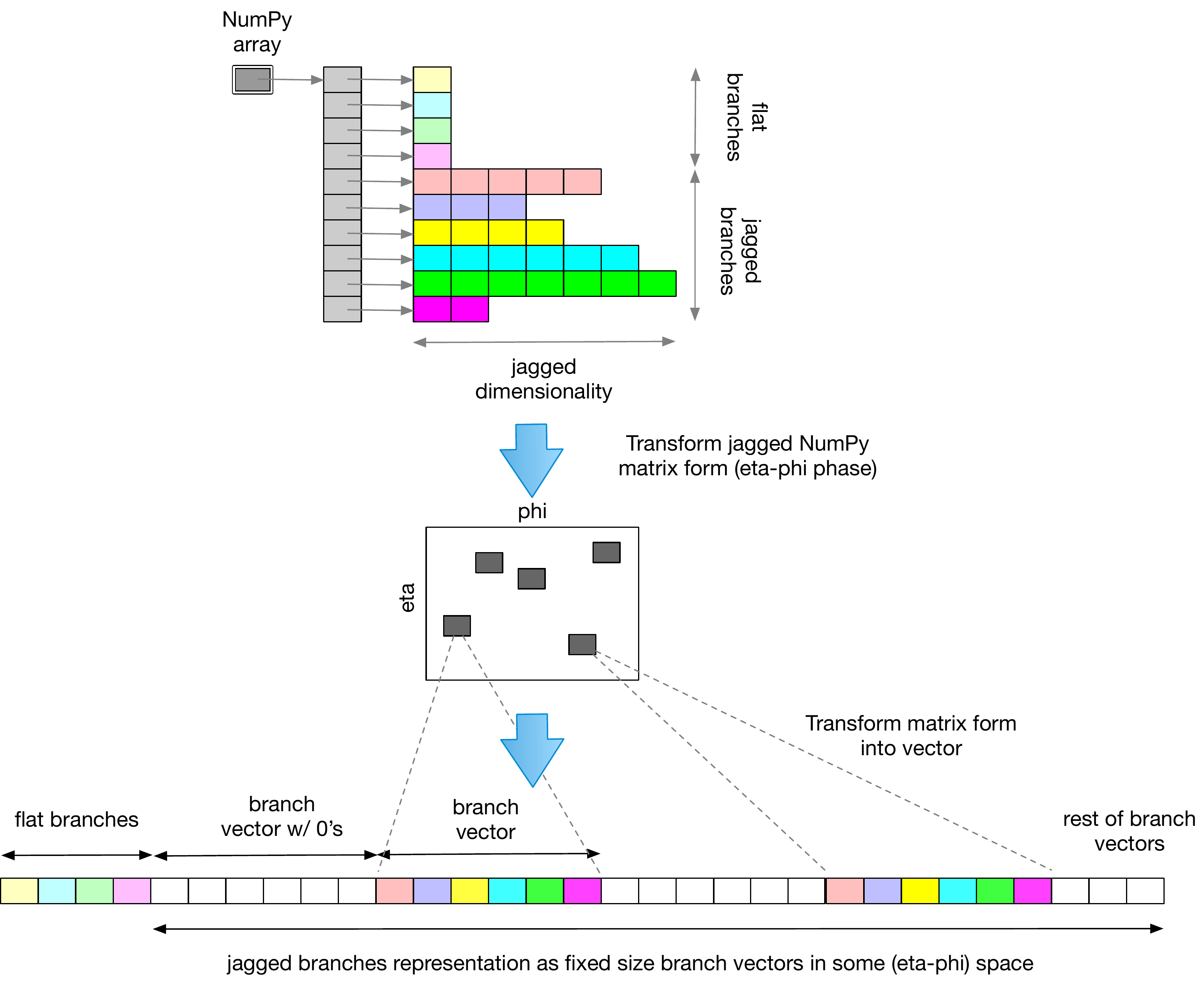}
    \caption{A matrix representation of Jagged Array into certain phase space, e.g. eta-phi \cite{MLaaS_Valentin}}
\label{fig:JaggedArray2Matrix}
\end{figure*}

This layer transforms HEP ROOT data presented by the Data Streaming Layer as Jagged Array
into a flat data-format used by the application \cite{MLCWP,existing_workflows}. 
The Jagged Array (see Fig. \ref{fig:JaggedArray}) is a compact representation
of variable size event data produced in HEP experiments.
The HEP tree-based data representation is optimized for data storage but it is not directly
suitable for ML frameworks. Therefore a certain data transformation is required to
feed tree-based data structures into the ML framework as a flat data structure.
We explored two possible transformations: a vector representation with padded values
(see Fig. \ref{fig:JaggedArray2Vector}) and a matrix representation of data into the phase
space of user choice (see Fig. \ref{fig:JaggedArray2Matrix}).

The HEP events have different dimensionality across event attributes.  For
instance, a single event may have a different number of physics particles.
Therefore, proper care should be done to flatten and padding ROOT events
in the Jagged Array representation. For that, we use a two-passes procedure. In
the first pass across all the events\footnote{Even though this procedure may not be
feasible at Peta-Byte scale, it can be easily replaced by studying various
Monte-Carlo distributions of such events to find attribute's boundaries.} we determine the dimensionality of
each attribute and its min/max values. In the second pass we map Jagged
Array attributes into a single vector representation with proper size and
padding (see Fig. \ref{fig:JaggedArray2Vector}).  In addition, we provide a
proper normalization of each attribute during this phase\footnote{This layer
can be easily abstracted as a Python decorator to allow multiple implementations
of normalization procedure.}.
\begin{figure*}[!t]
\centering
    \includegraphics[width=0.8\textwidth]{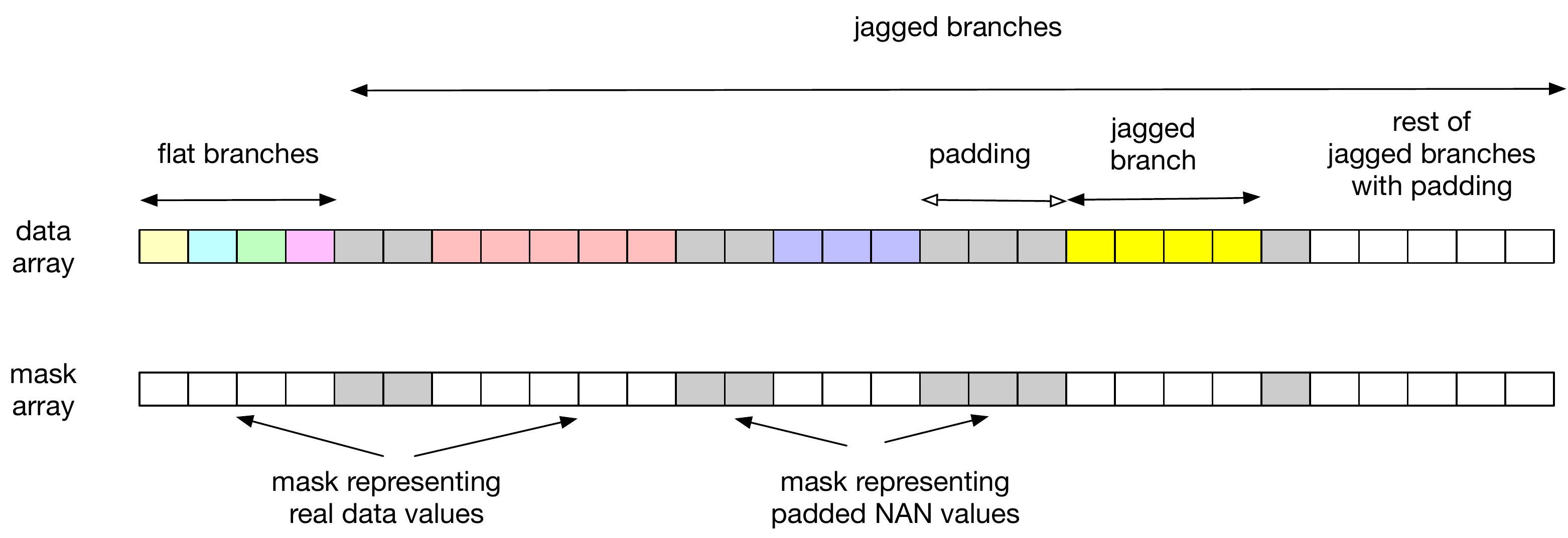}
    \caption{A vector representation of Jagged Array along with corresponding mask vector \cite{MLaaS_Valentin}}
\label{fig:JaggedArrayMask}
\end{figure*}
We also keep a separate masking vector (see Fig. \ref{fig:JaggedArrayMask})
to distinguish assigned padded (e.g. NaN or zeros) values from the real
values of the attributes. This may be important in certain kinds of Neural
Networks, e.g. AutoEncoders (AE) \cite{MLBook}
where the location of padded values in the input
vector can be used in the decoding phase.

A matrix representation can be obtained from a Jagged Array (see Fig.
\ref{fig:JaggedArray2Matrix}).  For example, the spatial coordinates or the
attribute components are often part of HEP datasets, and therefore it can be used
for this mapping.  This approach can resolve the ambiguity\footnote{For very
large datasets we may use Monte-Carlo distributions to determine
the finite size of the certain attributes and cut them off at a certain level, and, 
therefore, reject rare events which may exceed this threshold.} of vector
representation (in terms of dimensionality choice) but it has its own problem
with the choice of granularity of space matrix. For example, in
the simplest case, a 2D matrix representation\footnote{In the general case the
matrix representation can have any number of dimensions.} (see Fig.
\ref{fig:JaggedArray2Matrix}) can be used in some X-Y phase space. In this case,
this matrix represents an image where X and Y refer to an arbitrary pair of
attributes. But the cell size of this image is not known a-priory.
A choice of cell size may introduce a collision problem within an event, e.g.
different particles may have values of (X,Y) pair within the same cell.
Such ambiguity may be easily resolved either by increasing matrix granularity or
using another phase space, e.g. via higher dimensions of the cell
space.  But such changes will increase the sparsity of matrix representation 
and the matrix size, and
therefore will require more computing resources at the training time.

Below we provide details of the MLaaS4HEP workflow used in the
Data Streaming and Data Training layers using a vector representation
for the results presented in Sect. \ref{real-case}.

\subsection{ML training workflow implementation\label{workflow}}
We implemented the Data Streaming and Data Training layers using Python
programming language and we made them available in the MLaaS4HEP repository
\cite{MLaaS4HEP} under MIT license. The Data Training Layer was abstracted to support any kind of
Python-based ML frameworks: TensorFlow, PyTorch, and others\footnote{In
all our tests we used Keras and PyTorch frameworks to define our ML models.}.

We used two parameters to control the data flow within the framework.  The 
$chunk\ size$ parameter controls a chunk of data read by the Data Streaming
Layer from local or remote storage. And, the $batch\ size$ parameter
defines the number of events used by the underlying ML framework in each
training cycle. Therefore, further, we refer to chunk as a set of
events read by the Data Streaming Layer while batches as a set of events
used by the ML training loop.

\begin{figure*}[!t]
\centering
    \includegraphics[width=1\textwidth]{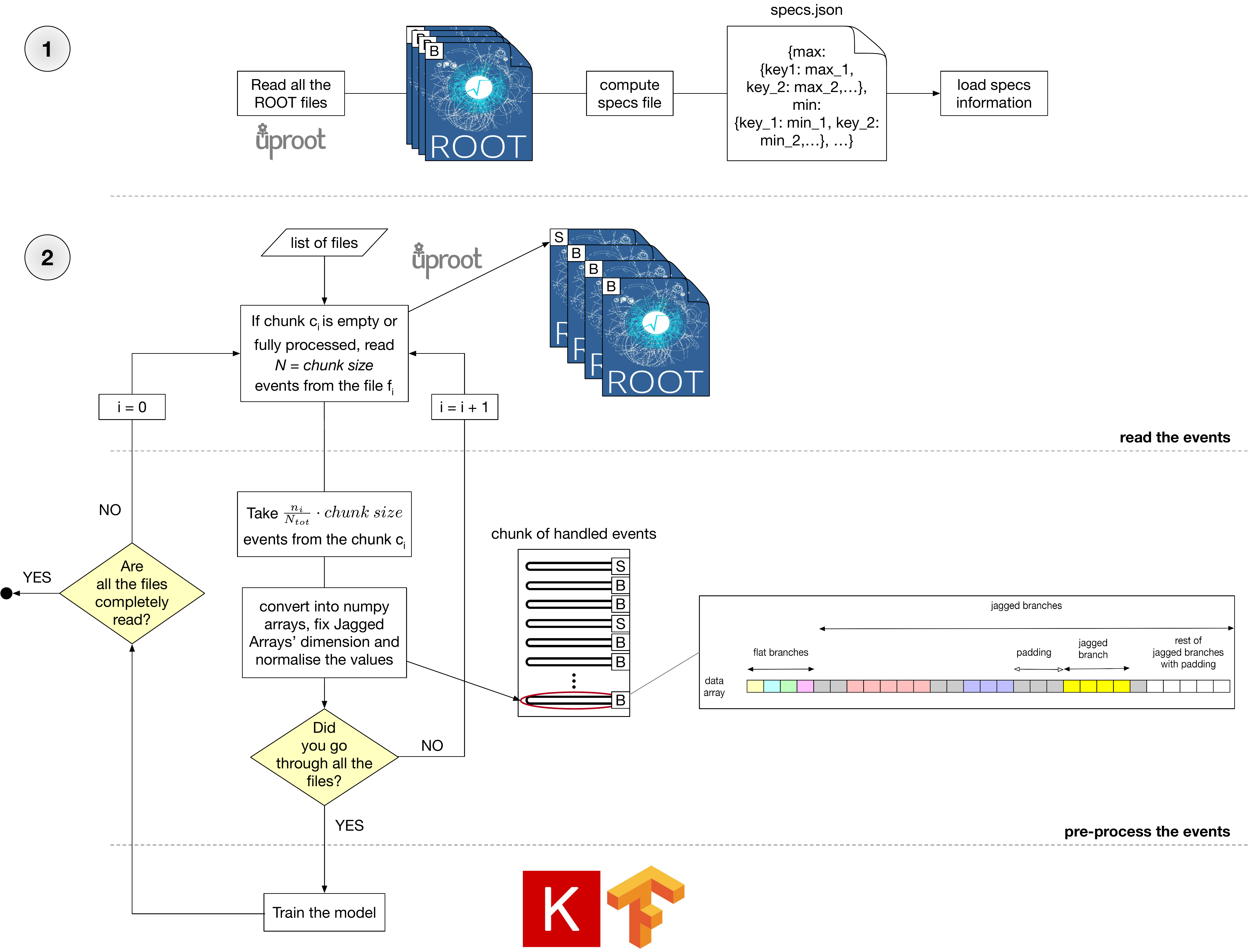}
    \caption{Schematic representation of the steps performed in the MLaaS4HEP pipeline, in particular those inside the Streaming and Training layers (see text for details)}
\label{fig:update_model}
\end{figure*}

In order to train the ML models defined by the user code (provided externally) the MLaaS4HEP
framework uses proper data chunks with the same proportion of events presented in 
the ROOT files. The schematic of the data flow
used in the Data Streaming and Data Training layers is shown in Fig.
\ref{fig:update_model}.

The first pass (denoted by $\textcircled{1}$ in Fig. \ref{fig:update_model})
represents the reading part of the MLaaS4HEP pipeline to create a specs file.
This part is performed by reading all the ROOT files in chunks (which size is
fixed a priori by the user) so that the information stored in the specs file is 
updated chunk by chunk. The specs file contains all the information about the 
ROOT files: the dimension of Jagged branches, the minimum and the maximum for 
each branch, and the number of events for each ROOT file\footnote{
Once the specs file is produced, either through the aforementioned procedure or
by studying Monte-Carlo distributions (for large datasets) to determine attribute dimensions and
their min/max values, it can be reused for all files from the given dataset
during the ML training phase.}.

The second part of the 
flowchart shown as $\textcircled{2}$ represents the following logic of the ML training phase.
%At the beginning of the cycle, the chunk $c_i$ is empty, and the events are not
%read yet.  In the first loop of the cycle we read $N$ events (where $N$ equal
%to the $chunk\ size$) from the file $f_i$ that are stored into the chunk $c_i$. Then, we
%use only a portion of those events to create the chunk used to train the ML
%model. In the next cycle, we read the right amount of data from the chunk $c_i$ to
%continue ML workflow. After several ML training loops we completed reading
%the events from the chunk $c_i$, and, therefore, we read another $N$ events that
%we store in chunk $c_i$, and so on.  In conclusion, we read from the file $f_i$ the
%number of events equal to the chunk size only when our chunk of events to be
%processed is empty (e.g. first time of the loop), or only we completed to read
%and process this chunk of events in ML training loops.
%Then, $n_i/N_{tot}\cdot chunk\ size$ events are taken from it, where $n_i$ is the number of
%the events of the i-th file and $N_{tot}$ is the whole amount of events of all
%the files. These events are converted into Numpy arrays, with a necessary fix
In the first loop of the cycle, when the events are not read yet, we read 
$N$ events (where $N$ is equal to the $chunk\ size$) from the i-th file $f_i$ 
that we store into the i-th chunk $c_i$.
Then $n_i/N_{tot}\cdot chunk\ size$ events are taken from it, where $n_i$ is the number of
the events of the file $f_i$ and $N_{tot}$ is the whole amount of events of all
the files. These events are converted into Numpy arrays, with the necessary fix
of the Jagged Arrays dimensions and normalization of the values. This part is performed
thanks to the information contained in the specs file computed in step
$\textcircled{1}$. The reading of the events and their pre-processing is performed for all the files $f_i$. 
After having created a chunk of $N$ events properly mixed from 
the different files, the events are used to train the ML model. The training 
phase is performed using batches of data taken from the created chunk,
and run for a certain number of epochs. The batch size and the number of 
epochs are fixed a priori by the user. 
Then we come back at the beginning of the cycle, and if all the events stored in the chunk $c_i$
have been already read, we read $N$ events from the file $f_i$, otherwise we read the proper
amount of events ($n_i/N_{tot}\cdot chunk\ size$) from the chunk $c_i$. Subsequently, the
events are pre-processed. The part of reading and pre-processing is performed for all the files $f_i$, in order to
create the proper data chunk used to train the ML model. Finally, the model is trained.
If the files are not completely read the entire pipeline is restarted from the beginning 
of point $\textcircled{2}$ until all events are read, creating at each cycle a new 
chunk of events that is used to train the ML model. At the end of this cycle, all 
the events contained in all files are read and the training process of the model 
is completed, producing a model that can be used in physics analysis.

\subsection{Data Inference Layer}\label{Inference}
A data inference layer can be implemented in a variety of ways.
It can be either tightly integrated
with application frameworks (for example both CMS and ATLAS experiments followed
this approach in their CMSSW-DNN \cite{CMSSWDNN} and LTNN \cite{ATLASLNN}
solutions respectively) or it can be developed as a Service (aaS) solution. The former
has the advantage of reducing latency of the inference step per processing
event, but the latter can be easily generalized and become independent from 
internal infrastructure. For instance, it can be easily integrated into
cloud platforms, it can be used as a repository of pre-trained models, and also
serve models across experiment boundaries. However, the speed of the data inference
layer, i.e. throughput of serving predictions, can vary based on the chosen technology.
A choice of HTTP protocol guarantees easy adaptation, while gRPC protocol
can provide the best performance but will require dedicated clients.
We decided to implement the Data Inference Layer as a
TensorFlow as a Service architecture \cite{TFaaS} based on
HTTP protocol\footnote{The code is available in the TFaaS repository \cite{TFaaS} under MIT license.}.

We evaluated several ML frameworks and we decided to use TensorFlow graphs
\cite{TF} for the inference phase. The TF model represents a computational
graph in a static form, i.e. mathematical computations, graph edges, and data
flow are well-defined at run time. Reading TF model can be done in different
programming languages thanks to the support of APIs provided by the TF library.
Moreover, the TF graphs are very well optimized for GPUs and TPUs. We opted for
the Go programming language \cite{GoLang} to implement the inference part of the
MLaaS4HEP framework based on the following factors: the Go language natively
supports concurrency via \textit{goroutines} and \textit{channels}; it is the
language developed and used by Google, and it is very well integrated with the TF
library; it provides a final static executable which significantly simplifies
its deployment on-premises and to various (cloud) service providers.  We also
opted out in favor of the REST interface. Clients may upload their TF models
to the server and use it for their inference needs via the same interface. Both
Python and C++ clients were developed on top of the REST APIs (end-points) and
other clients can be easily developed thanks to HTTP protocol.  The TFaaS
framework can be used outside of HEP to serve any kind of TF-based models
uploaded to TFaaS service via HTTP protocol\footnote{For instance, we tested the 
TFaaS functionality using non-HEP models such as image recognition ML models.}.

\subsection{MLaaS4HEP: proof-of-concept prototype}\label{Prototype}
When all layers of the MLaaS4HEP framework were developed, we successfully
tested a working prototype of the system by using ROOT files accessible through
XrootD servers. The data were read in chunks of 1k events, where the single
chunk was approximately 4 MB in size. We tested this prototype on a local
machine as well as successfully deployed it on a GPU node. To further validate
the MLaaS4HEP framework we decided to apply it to a real physics analysis, see
Sect. \ref{real-case}, where we explored local and remote data access,
usage of different data chunks, random access to files, etc. All details
can be found in the next section.

\section{Real case scenario}\label{real-case}
In order to validate the MLaaS4HEP approach, we decided to test the
infrastructure on a real physics use-case. This allowed us to test the
performances of the MLaaS4HEP framework, and validate its results from the
physics point of view. We decided to use the $t\bar{t}$ Higgs analysis
($t\bar{t}H(b\bar{b})$) in the boosted, all-hadronic final state \cite{ttH_CMS,ttH_allJets,ttH_CMS_note} 
due to affinity with the analysis group. In the following sub-sections we discuss:
\begin{itemize}
\item the $t\bar{t}H(b\bar{b})$ all-hadronic analysis strategy (Sect. \ref{physics});
\item MLaaS4HEP validation (Sect. \ref{validation});
\item MLaaS4HEP performance results using the physics use-case (Sect. \ref{performance});
\item MLaaS4HEP projected performance (Sect. \ref{projection});
\item TFaaS performance results (Sect. \ref{TFaaSPerformance}).
\end{itemize}

\subsection{$t\bar{t}H(b\bar{b})$ all-hadronic analysis strategy}\label{physics}
The Higgs boson is considered the most relevant discovery of the last few years
in High Energy Physics. After almost fifty years from its prediction, it was
discovered by the ATLAS and CMS collaborations in 2012 at the CERN Large-Hadron
Collider (LHC) \cite{Higgs_ATLAS,Higgs_CMS}. Since then, many analyses have
been performed in order to measure its properties with higher precision.

\begin{figure}[h]
\centering
    \includegraphics[width=0.35\textwidth]{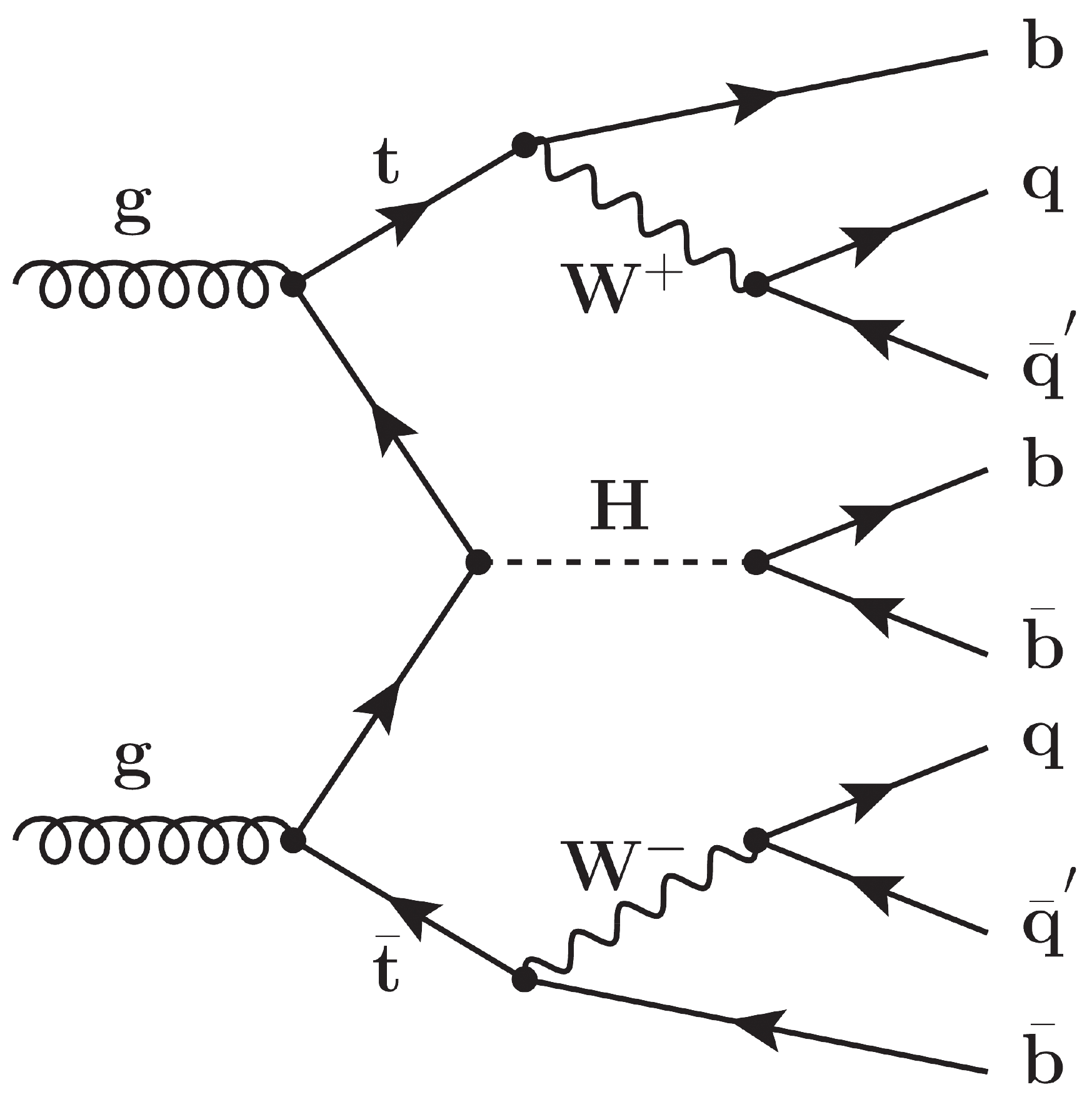}
    \caption{Feynman diagram for the $t\bar{t}H(b\bar{b})$ decay}
\label{fig:decay_plot}
\end{figure}

In the Standard Model framework, the Higgs boson is predicted to couple with
fermions via Yukawa-like interaction, which gives the mass to fermions
proportionally to the coupling. The heaviest top quark is responsible for
coupling to the Higgs boson.  Direct measurement of the top-Higgs coupling
exploits tree-level processes. The $t\bar{t}H$ production plays an important 
role in the study of the top-Higgs Yukawa coupling, as other production 
mechanisms (such as gluon-gluon fusion) involve loop-level diagrams in which 
contributions from Beyond Standard Model (BSM) physics could enter the loops 
unnoticed. The highest branching ratio ($\approx$25\%) is represented by the
all-hadronic decay channel with $H(b\bar{b})$ and all-hadronic $t\bar{t}$.
The W bosons produced by the $t\bar{t}$ pair decay into a pair of light 
quarks while the Higgs boson decays into a $b\bar{b}$ pair (see Fig.
\ref{fig:decay_plot}). In the final state, there are at least eight partons (more
might arise from the initial and final state radiation) where four of them are
bottom (b) quarks. Despite the highest branching ratio, the all-jets final
state is very challenging. It is dominated by the large QCD multi-jet
production at LHC, and there are large uncertainties in this channel due to the
presence of many jets. At the same time, it represents the unique possibility
to fully reconstruct the $t\bar{t}H$ as all decay products are observable.

At the 13 TeV center-of-mass energy, top quarks with a very high $p_{T}$ can be 
produced via $t\bar{t}H$. If their Lorentz boost is sufficiently high, their 
decay products are very collimated into a single, wide jet, named boosted jet. 
In particular, we are interested in the $t\bar{t}H(b\bar{b})$ analysis with 
all-jets final state
where at least one of the jets of the final state is a boosted jet, and 
where the Higgs boson decays in a pair of well resolved jets identified as 
a result of the hadronization of bottom quarks.

For identification of the $t\bar{t}H(b\bar{b})$ events containing a resolved-Higgs 
decay a Machine Learning model based on Boosted Decision Tree (BDT) was used by 
the CMS Higgs Physics Analysis Group (HIG PAG) \cite{ttH_CMS_note} 
and the training was done within TMVA \cite{TMVA} framework.
The Monte Carlo simulation provides events used for training, where events are
selected among the $t\bar{t}H$ sample and the two dominant
background samples, namely QCD and $t\bar{t}$, respectively.
The $t\bar{t}H$ events with the 
resolved Higgs-boson matching to the system of two b-tagged jets are considered 
as signal events. On the contrary, unmatched $t\bar{t}H$ events, and all the 
QCD and $t\bar{t}$ events are considered  as background events. Both signal 
and background events are required to pass some selection criteria, such as to have 
at least a boosted jet, to contain no leptons, to pass the signal trigger, etc. 
This selection is aimed to select boosted, all-jets-like events.

\subsection{MLaaS4HEP validation \label{validation}}
In order to validate the MLaaS4HEP functionality against standard BDT-based procedure,
we decided to use a set of ROOT files from the resolved-Higgs analysis discussed in Sect. \ref{physics}.
The goal of this exercise was to demonstrate that the MLaaS4HEP framework can
provide a valuable alternative and deliver comparable results with
respect to the traditional analysis based on a pre-defined set of metrics.
For our purposes, we decided to use a generic ML model and compare the
results obtained inside and outside MLaaS4HEP.
In particular, we explored the following approaches:
\begin{itemize}
    \item use MLaaS4HEP to read and normalize events, and to train the ML model;
    \item use MLaaS4HEP to read and normalize events, and use a Jupyter notebook to perform the training of the ML model outside MLaaS4HEP;
    \item use a Jupyter notebook to perform the entire pipeline without using MLaaS4HEP.
\end{itemize}

Initially, we performed the
analysis using the ROOT files that passed the selection criteria discussed in Sect. \ref{physics}.
The final dataset consisted of eight ROOT files containing background events, and one file
containing signal events. Each file had 27 branches, with 350k events in total,
and the total size of this dataset was 28 MB.  The ratio between the number of
signal events and background events was approximately 10.8\%. The dataset was 
split into three parts, 64\% for training, 16\% for validation, and 20\% for test 
purposes, respectively. In particular, we used a Keras sequential Neural Network 
with two hidden layers made by 128 and 64 neurons, and with a 0.5 dropout 
regularization between layers. Finally, we trained the model for 5 epochs with a
batch size of 100 events.

\begin{figure*}[t]
  \centering
  \begin{subfigure}[t]{0.55\linewidth}
    \includegraphics[width=\linewidth]{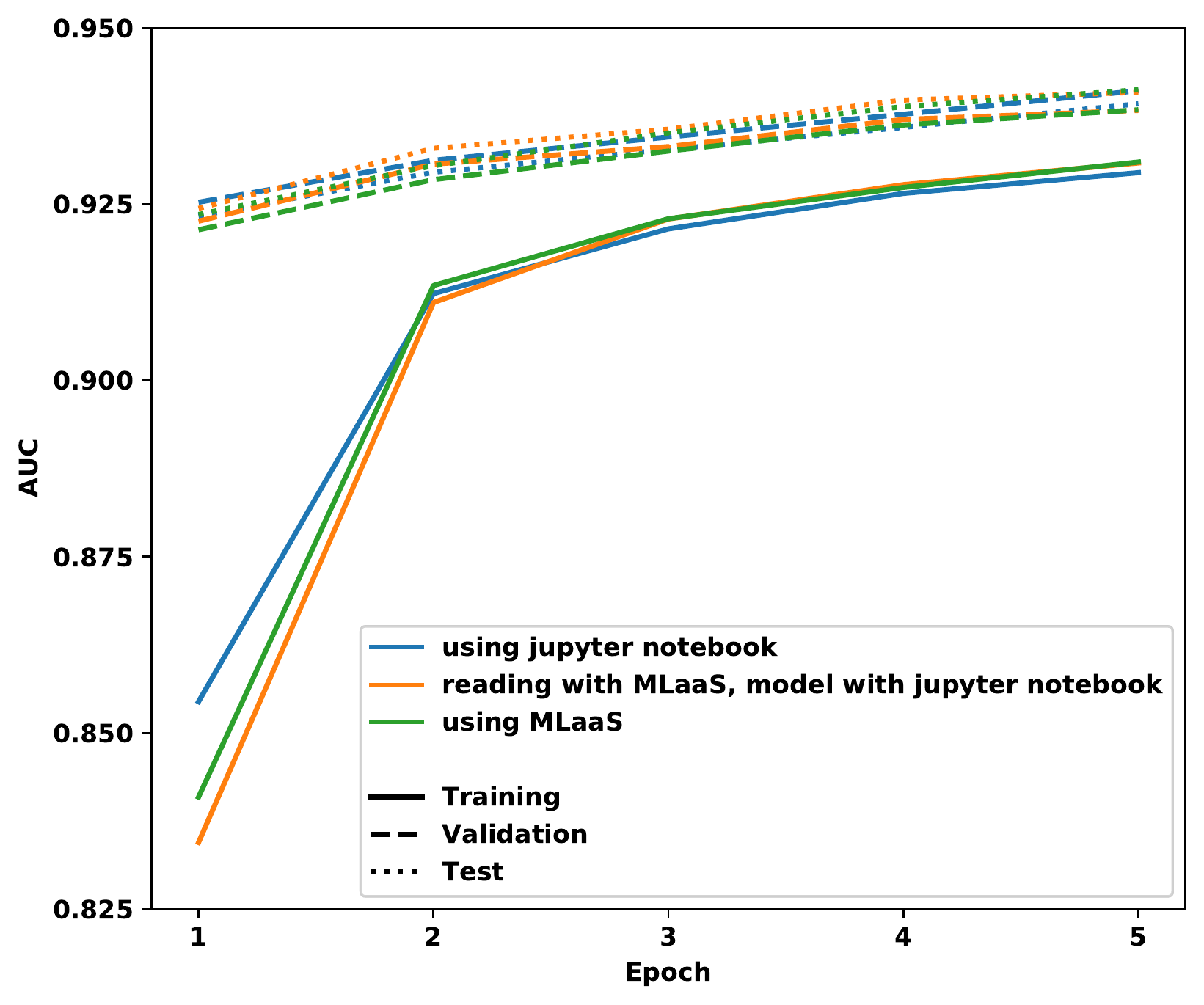}
  \end{subfigure}
    \caption{Comparison of the AUC score for the training, validation, and test
    set for three different cases: (i) using MLaaS4HEP to read and normalize
    events, and to train the ML model; (ii) using MLaaS4HEP to read and
    normalize events, and using a Jupyter notebook to perform the training of
    the ML model outside MLaaS4HEP; (iii) using a Jupyter notebook to perform
    the entire pipeline without using MLaaS4HEP}
  \label{fig:metrics_comparison}
\end{figure*}

The results of this exercise are shown in Fig. \ref{fig:metrics_comparison},
and demonstrate that different approaches have similar performance.
The AUC score of the generic ML model is found to be comparable with the BDT-based
analysis\footnote{Please note, our goal was to demonstrate that the MLaaS4HEP
approach provides similar results to the BDT-based analysis, but we did not
target to reproduce and/or match exact AUC numbers obtained in the standard
physics analysis.}.

\subsection{MLaaS4HEP performance\label{performance}}
In this section, we provide details of the MLaaS4HEP performance testing: the
scalability of the framework and its benchmarks using different storage layers.
For that purpose, we used all available ROOT files without any physics cuts.
This gave us a dataset with 28.5M events with 74
branches (22 flat and 52 Jagged), and a total size of about 10.1 GB.

We performed all tests running the MLaaS4HEP framework on
macOS, 2.2 GHz Intel Core i7 dual-core, 8 GB of RAM, and on CentOS 7 Linux, 4
VCPU Intel Core Processor Haswell 2.4 GHz, 7.3 GB of RAM CERN Virtual Machine.
The ROOT files are read from three data-centers: Bologna (BO), Pisa (PI), and Bari (BA).

Table \ref{tab:reading_all_files} summarizes the I/O numbers we obtained in the
first step of the MLaaS4HEP pipeline ($\textcircled{1}$ in Fig.
\ref{fig:update_model}) using various setups and a chunk size of 100k events.
It provides the values of time spent for reading the files, the time spent
for computing specs values, 
the total time spent for completing the step $\textcircled{1}$,
and the event throughput for the reading and specs computing step.

\begin{table*}[ht]
\begin{center}
\begin{tabular}{ |l|c|c|c|c| } 
 \hline
 & reading time & specs comp. time & time to complete & event throughput for \\
 & (s) & (s) & step $\textcircled{1}$ (s) & reading + specs comp. (evts/s) \\
 \hline
macOS with local files  & 1532 & 1031 & 2608 &  11006\\
 \hline
macOS with remote files (BO) & 4349 & 1007 & 5453  &  5265\\
 \hline
VM with local files & 1132 & 978 &  2153 & 13366 \\
 \hline
VM with remote files (BO) & 1919 & 1017 & 2994 & 9606 \\
 \hline
VM with remote files (BA) & 2136 & 988 & 3193 & 9027 \\
 \hline
VM with remote files (PI) & 2114 & 996 & 3171 & 9067 \\
 \hline
\end{tabular}
    \caption{Performances of reading and specs computing phase with chunk size
    fixed to 100k events, using the macOS system and the CERN VM. In
    local storage cases, the files are stored in a SSD 500 GB in the macOS case
    and in a Virtual Disk 52 GB in the CERN VM case, respectively. Moreover, BO, BA, and PI
    stand for various Italian storage facilities with different WAN
    configurations (see text for more details)}
\label{tab:reading_all_files}
\end{center}
\end{table*}

\begin{figure*}[ht]
\centering
    \includegraphics[width=0.85\textwidth]{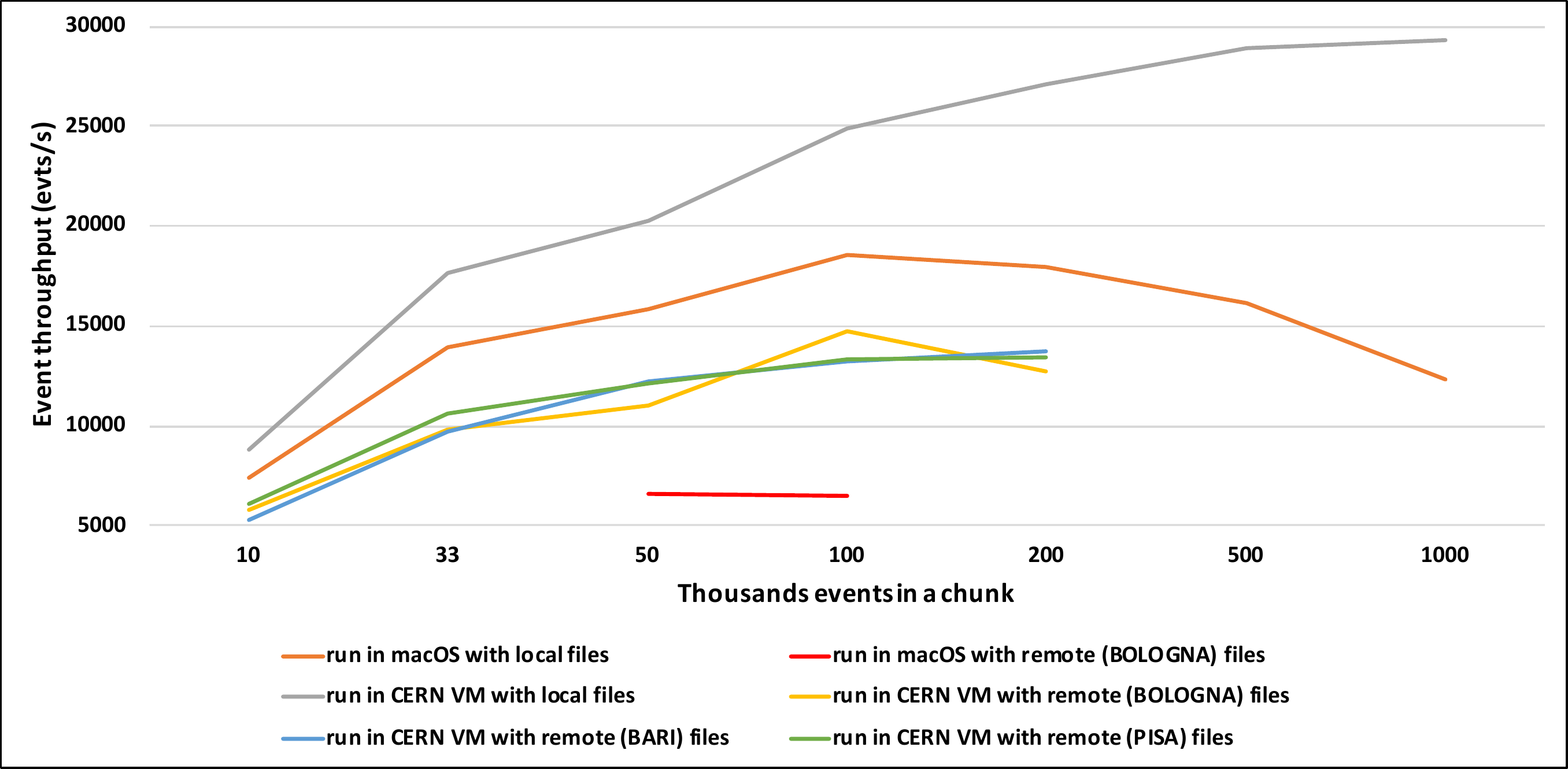}
    \caption{Event throughput for reading the data as a function of the chunk size for different trials}
\label{fig:reading_time}
\end{figure*}

In Fig. \ref{fig:reading_time} we show the event throughput for reading the data as a 
function of chunk size for different trials. In all cases, we find no significant 
peaks. The larger chunk sizes can lead to certain problems, as in the case of 
the CERN VMs, where we may reach a limitation of the underlying hardware, e.g. 
big memory footprint.

In the performance studies of the second step of the MLaaS4HEP pipeline 
($\textcircled{2}$ in Fig. \ref{fig:update_model}) we are interested in the data 
reading part, the data pre-processing step (which include data transformation), 
and the time spent in the MLaaS4HEP training step.

\begin{figure*}[ht]
\centering
    \includegraphics[width=0.85\textwidth]{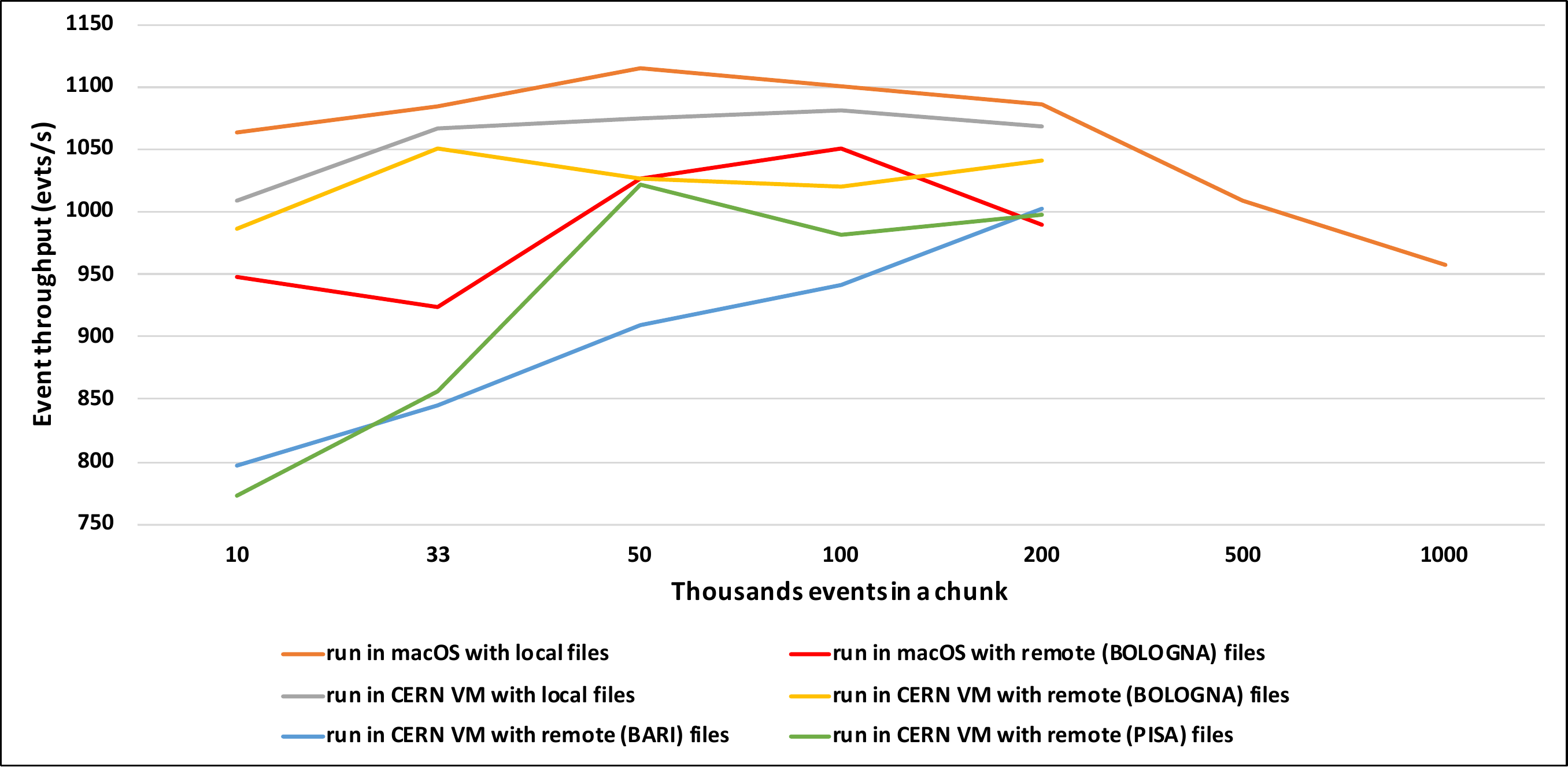}
    \caption{Event throughput for creating a chunk as a function of the chunk size for different trials}
\label{fig:creating_chunk_frequency}
\end{figure*}

As already mentioned in Sect. \ref{workflow}, there is a loop over files that allows building the chunk
used to train the ML model with the adequate proportion of the events. If the 
chunk that contains the events of the i-th ROOT file is empty or fully processed, a 
new chunk of events from the i-th file is read, and the time for reading 
is added to the whole time spent for creating the chunk (see Fig. \ref{fig:update_model}). 
In other words, the 
time spent for creating a chunk is made by the sum of $n$ reading actions, and 
of the time to pre-process the events. The event throughput for creating a single 
data chunk and the event throughput for pre-processing a 
single data chunk are reported in Table \ref{tab:create_chunk}.  In Fig. 
\ref{fig:creating_chunk_frequency} we show the event throughput for creating a 
chunk as a function of the chunk size for different trials.

We found that the time spent for creating a chunk was almost the same using
macOS or CERN VM, and similar using local or remote files. Obviously,
for remote files, the reading time increased consequently, and the time for
creating the chunk increased, but this difference was quite negligible.  
For instance, we spend around 90 seconds to create a chunk of 100k events, which 
translates into an event throughput of about 1.1k etvs/s as reported in Table 
\ref{tab:create_chunk}.

\begin{table*}[ht]
\begin{center}
\begin{tabular}{ |l|c|c| } 
 \hline
 & Event throughput for & Event throughput for\\
  & creating a chunk (evts/s) & pre-processing a chunk (evts/s)\\
 \hline
macOS with local files & 1101 & 1156\\
 \hline
macOS with remote files (BO) &  1051 & 1188\\
 \hline
VM with local files &  1081 & 1120\\
 \hline
VM with remote files (BO) & 1020 & 1080\\
 \hline
VM with remote files (BA) & 942 & 1081\\
 \hline
VM with remote files (PI) & 982 & 1060\\
 \hline
\end{tabular}
\caption{Event throughput for the chunk creation and for the pre-processing step with a chunk size of
    100k events computed as the ratio of the number of events over the
    time spent on chunk creation. The difference between the two steps is based
    on the reading part, i.e. the time for creating a chunk, as the sum of
    times for reading events from the ROOT files, and the time for
    the pre-processing step}
\label{tab:create_chunk}
\end{center}
\end{table*}

During the implementation of the MLaaS4HEP framework, we resolved few bottlenecks
with respect to the results obtained in \cite{MLaaS_Valentin}.
For example, we improved the reading time by a factor of 10. This
came from better handling of Jagged Arrays via flattening the event arrays
and computing of min/max values of each branch.
Moreover, we also obtained a factor of 2.8 improvements in the data
pre-processing step by using lists comprehensions
instead of loops within the event.
%\footnote{The underlying data
%transformation from uproot to Jagged arrays to lists (1.3 improvement factor),
%and data transformations, such as normalization, fixing dimensions, masking
%steps.}.
We found that MLaaS4HEP took about 53 seconds to pre-process 100k
events with 42\%, 44\%, and 10\% breakdown used for the normalization step,
fixing the dimensions, and creating the masking vectors, respectively.

In conclusion, for the presented physics use-case we found comparable results between
ML models inside and outside the MLaaS4HEP framework. Using 10 GB of data (approximately
28.5M events) we obtained the following results:
\begin{itemize}
    \item MLaaS4HEP framework is capable to work with local and remote files;
    \item its throughput reaches about 13.4k evts/s for reading local ROOT
        files (with specs computing), and about 9.6k evts/s for remote files;
    \item the throughput of pre-processing step is peaked at 12k evts/s.
\end{itemize}

\subsection{MLaaS4HEP performance projection\label{projection}}
Based on our studies presented in the previous section we found that
MLaaS4HEP takes about 35 minutes for the first 
step of the pipeline ($\textcircled{1}$ in Fig. \ref{fig:update_model}), 
and around 7 hours for the second step ($\textcircled{2}$ 
in Fig. \ref{fig:update_model}) to process a dataset of 10 GB of data (28.5M events).
%Therefore, we project that these
%values will increase by a factor $10^2$ and $10^5$ for datasets of TB 
%and PB scale, respectively. 
Therefore, we estimate that using the same hardware resources
the step $\textcircled{1}$ will take about
58 hours and 58k hours for datasets at TB and PB scale, and the time 
for step $\textcircled{2}$ will be around 719 hours and 719k hours, respectively.

At this stage, our goal was mainly to prove the feasibility 
of the MLaaS4HEP pipeline, and validate its usage within the context of a real
physics use-case rather than perform real ML training at TB/PB scale.
In Sect. \ref{future_directions} we discuss further improvements which can be done.

\subsection{TFaaS performance\label{TFaaSPerformance}}
The performance testing of the TFaaS service was done using a variety of
ML models, from simple image classification to the ML model developed and
discussed in Sect. \ref{validation}. In particular, we performed several
benchmarks using the TFaaS server running on CentOS 7 Linux, 16 cores, 30 GB of
RAM.  The benchmarks were done in two modes: using 1k calls with 100
concurrent clients and 5k calls with 200 concurrent clients.  We tested both
JSON and ProtoBuffer \cite{ProtoBuffer} data formats while sending and fetching
the data to/from the TFaaS server. In both cases, we achieved a throughput
of $\sim 500$ req/sec.  These numbers were obtained by serving mid-size
pre-trained model which consists of 1024x1024 hidden layers used in the physics
analysis discussed in Sect. \ref{physics}.  Even though a single TFaaS server
may not be as efficient as an integrated solution, it can be easily
horizontally scaled, e.g. using Kubernetes or other cluster orchestrated solutions, and may
provide the desired throughput for concurrent clients. It also decouples the
application layer/framework from the inference phase which can be easily
integrated into any existing infrastructure by using the HTTP protocol to TFaaS
server for inference results.
We foresee that it can be useful in a 
variety of use-cases such as quick evaluation of ML models in physics analysis,
or online applications where new models can be built periodically.
The TFaaS implementation allows to use itself as a repository of ML pre-trained
models, and it can be a valuable component in the agile ML development cycle of any
group, from small physics analysis group(s) to cross-experiment collaborations.

\section{Future directions}\label{Improvements}
\label{future_directions}
In the previous section, we discussed the usage of MLaaS4HEP in the scope of a real HEP physics
analysis. We found the following:
\begin{itemize}
    \item the usage of MLaaS4HEP is transparent to the chosen HEP dataset, i.e.
        data can be read locally or from remote storage;
    \item the discussed architecture is HEP experiment agnostic and can be
        used with any existing ML (Python-based) framework as well as easily
        integrated into existing infrastructure;
    \item the data can be read in chunks from remote storage,
        and this allows continuous ML training over the large datasets,
        and further parallelization.
\end{itemize}
These observations open up a possibility to train ML models over large datasets,
potentially at Peta-Byte scale, while using existing Python-based open-source
ML frameworks. Therefore, we foresee that the Machine Learning as a Service
approach can be widely applicable in HEP.
For example, future directions of this work might include the exploitation of
this architecture to streamline the access to cloud and HPC resources for
training and inference tasks. It can represent an attractive option to open up
HPC resources for large scale ML training in HEP along with required security
measurements, resource provisioning, and remote data access to WLCG sites.
To move in this direction additional work will be required.
Below, we discuss a possible set of improvements that can be explored.

\subsection{Data Streaming Layer}
To improve the Data Streaming Layer a multi-threaded I/O layer can be
implemented. This can be achieved by wrapping up the data reader code-base into
a service that will deliver the data chunks in parallel upon requests from
the upstream layer. In addition, the chunks can be pre-fetched from XrootD servers
into a local cache to improve the I/O throughput. In particular, there are
several R\&D's underways to demonstrate intelligence smart caching
\cite{SmartCache} for Dynamic On-Demand Analysis Service (DODAS) at computer
centers, such as HPC, national Tier centers, etc. Such a DODAS facility can
reduce the time spent on the Data Streaming Layer by pre-fetching ROOT files into
local cache and use them for ML training.

\subsection{Data Training Layer}
If data I/O parallelism can be achieved,
further improvements can be made via implementation of
distributed training \cite{distributed_training}. There are several R\&D developments in
this direction, from adapting the Dask Python framework \cite{Dask}, 
to using MPI-Based Python framework for distributed training \cite{MPIKeras}, or using
MLflow framework \cite{MLflow} on an HDFS+Spark infrastructure, which explores
both task and data parallelism approaches.

The current landscape of ML frameworks is changing rapidly, and we should be
adjusting MLaaS4HEP to existing and future ML framework and innovations. For
instance, Open Network Exchange Format \cite{onnxai} opens up the door to
migration of models from one framework into another. This may open up a
possibility to use MLaaS4HEP for the next generation of Open-Source ML frameworks
and ensure that end-users will not be locked into a particular one. For instance,
we are working on the automatic transformation of PyTorch \cite{PyTorch} and fast.ai
\cite{fastai} models into TensorFlow which later can be uploaded and used through
TFaaS service \cite{TFaaS}.

As discussed in Sect. \ref{Training} there are different approaches to feed
Jagged Arrays into ML framework and R\&D in this direction is in progress.  For
instance, for AutoEncoder models, the vector representation with padded
values should always keep around a cast vector which later can be used to
decode back the vector representation of the data back to Jagged Array or ROOT
TTree data-structures. We also would like to explore matrix representation of
Jagged Array data and see if it can be applied to certain types of use-cases,
e.g. in calorimetry or tracking where image representation of the objects can
be used.

\subsection{Data Inference Layer}
On the inference side, several approaches can be used. As discussed above, the
TFaaS \cite{TFaaS} throughput can be further improved by switching from HTTP
to a gRPC-based solution such as SONIC \cite{SONIC} which can provide a fast
inference layer based on FPGAs and GPUs-based infrastructures.

The current implementation of TFaaS can be used as a repository of pre-trained models which can be
easily shared across experiment boundaries or domains thanks to serving ML
models via HTTP protocol. For instance, the current implementation of TFaaS
allows visual inspection of uploaded models, versioning, tagging, etc.  We
foresee the next logical step is towards a repository of pre-trained models with
flexible search capabilities, extended model tagging, and versioning. This can
be achieved by providing a dedicated service for ML models with proper
meta-data description. For instance, such meta-data can capture model parameters,
details of used software, releases, data input, and performance output. With a
proper search engine in place, users may search for available ML models
related to their use-case.

\subsection{MLaaS4HEP services}
The proposed architecture allows us to develop and deploy training and inference
layers as independent services. The separate resource providers can be used
and dynamically scaled if necessary, e.g.  GPUs/TPUs can be provisioned
on-demand using the commercial cloud(s) for training purposes of specific
models, while inference TFaaS service can reside elsewhere, e.g. on a dedicated
Kubernetes cluster at some computer center.  For instance, the continuous
training of complex DL models would be possible when data produced by the
experiment will be placed on WLCG sites. The training service will receive
a set of notifications about newly available data, and re-train specific
model(s). When a new ML model is ready it can be easily pushed to TFaaS and be
available for end-users immediately without any intervention on the existing
infrastructure as part of CD/CI (Continuous Development and Continuous
Integration) workflows.  The TFaaS can be further adapted to use FPGAs to
speed up the inference phase.  We foresee that such an approach may be more
flexible and cost-effective for HEP experiments in the HL-LHC era.  As such, we
plan to perform additional R\&D studies in this direction and evaluate further
MLaaS4HEP services using available resources.

\section{Summary\label{summary}}
In this paper, we presented a modern approach to train HEP ML models using the
native ROOT data-format either from local or remote storage.
The MLaaS4HEP consists of three layers: the
Data Streaming and Data Training layers as part of the MLaaS4HEP framework
\cite{MLaaS4HEP}, and the Data Inference Layer implemented in the TFaaS framework based on the TensorFlow library.
All three layers are implemented as independent components. The
Data Streaming Layer relies on the uproot library for reading data from ROOT
files (local or remote) and yielding NumPy (Jagged) arrays. The Data
Training Layer transforms the input Jagged Array into
a vector representation and passes it into the ML framework provided by the
user. Finally, the Data Inference Layer was implemented as an independent HTTP service.

The flexible architecture we implemented allows performing ML training over
a large set of distributed HEP ROOT data without physically downloading data into
local storage. We demonstrated that such architecture is capable of
reading local and distributed datasets, available via XrootD protocol on WLCG
infrastructure. We validate the MLaaS4HEP architecture using an official CMS
$t\bar{t}$ Higgs analysis ($t\bar{t}H(bb)$) in the boosted, all-hadronic final
state, and we obtained comparable ML model performance with respect to a
traditional physics analysis based on data extraction from ROOT files into
custom Ntuples.

\begin{acknowledgement}
This work was done as a part of the CMS experiment R\&D program. We would like to
thank Jim Pivarski for his numerous and helpful discussions, and hard work
on uproot (and many other) packages which open up a possibility to
work on the MLaaS4HEP implementation. We would like to thank Fabio Iemmi for the
helpful discussions we had on the aspects of the physics use-case.
\end{acknowledgement}

% Authors must disclose all relationships or interests that 
% could have direct or potential influence or impart bias on 
% the work: 
%
\section{Declarations}
\section*{Funding}
Not applicable
 \section*{Conflict of interest}
 The authors declare that they have no conflict of interest.
 \section*{Availability of data and material}
 The details about datasets used in current study are
 available from the corresponding author on reasonable request.
 The availability of CMS dataset itself is a subject of CMS policy.
 \section*{Code availability}
 The code is available at \cite{MLaaS4HEP,TFaaS} under MIT license.

% BibTeX users please use one of
%\bibliographystyle{plain}
%\bibliographystyle{spbasic}      % basic style, author-year citations
%\bibliographystyle{spmpsci}      % mathematics and physical sciences
%\bibliographystyle{spphys}       % APS-like style for physics
%\bibliography{}   % name your BibTeX data base

% Non-BibTeX users please use

\end{document}